\documentclass{aa}  
\usepackage{graphicx}
\usepackage{natbib}
\usepackage{txfonts}
\def\hhhop{H$_3$O$^+$}
\def\hho{H$_2$O}
\def\hhoe{H$_2^{18}$O}
\def\hh{H$_2$}
\def\nn{N$_2$}
\def\oo{O$_2$}
\def\hhhp{H$_3^+$}
\def\nnhp{N$_2$H$^+$}
\def\hcop{HCO$^+$}
\def\dcop{DCO$^+$}

\def\zcr{$\zeta_{\rm CR}$}
\def\pow#1#2{#1$\times$10$^{#2}$}
\def\smm{(sub-)mil\-li\-me\-ter}
\def\opr{$o$/$p$}
\def\gtsim{{_>\atop{^\sim}}}
\def\ltsim{{_<\atop{^\sim}}}
\def\kkms{K\,km~s$^{-1}$}
\def\kms{km~s$^{-1}$}
\def\tmb{$T_{\rm mb}$}
\def\txc{$T_{\rm ex}$}

\def\vlsr{V$_{\rm LSR}$}
\def\ccm{cm$^{-3}$}
\def\scm{cm$^{-2}$}

\def\rs{s$^{-1}$}
\def\dv{$\Delta$\textit{V}}
\def\mic{$\mu$m}
\begin{document}
\title{APEX mapping of \hhhop\ in the Sgr~B2 region}

\author{F.F.S.~van der Tak\inst{1,2} \and A.~Belloche\inst{1} \and P.~Schilke\inst{1} \and
        R. G\"usten\inst{1} \and S. Philipp\inst{1} \and C. Comito\inst{1} \and P. Bergman\inst{3} \and L.--\AA. Nyman\inst{3}}

\institute{Max-Planck-Institut f\"ur Radioastronomie, Auf dem H\"ugel 69, 53121 Bonn, Germany 
      \and National Institute for Space Research (SRON), Postbus 800, 9700 AV Groningen, The Netherlands; \email{vdtak@sron.rug.nl} 
      \and European Southern Observatory, Casilla 19001, Santiago, Chile}

\date{Received 27 March 2006 / Accepted 22 May 2006}

\titlerunning{APEX mapping of \hhhop\ in the Sgr~B2 region}
\authorrunning{F.F.S. van der Tak et al.}

% \abstract{}{}{}{}{} 
% 5 {} token are mandatory
 
  \abstract
   {The cosmic-ray ionization rate \zcr\ of dense molecular clouds
   is a key parameter for their dynamics and chemistry.}
   {Variations of \zcr\ are well established, but it is unclear if
  these are related to source column density or to Galactic location.}
   {Using the APEX telescope, we have mapped the 364~GHz line of \hhhop\
    in the Sgr~B2 region and observed the 307~GHz line at selected
    positions. With the IRAM 30-m telescope we have observed the
    \hhoe\ 203\,GHz line at the same positions.}
  {Strong \hhhop\ emission is detected over a $\sim$3$\times$2\,pc region,
    indicating an \hhhop\ column density of 10$^{15}$--10$^{16}$\,\scm\ in an
    18$''$ beam.  The \hhhop\ abundance of $\sim$\pow{3}{-9} and \hhhop/\hho\ 
    ratio of $\sim$1/50 in the Sgr~B2 envelope are consistent with
    models with \zcr$\sim$\pow{4}{-16}\,s$^{-1}$, 3$\times$ lower than derived
    from \hhhp\ observations toward Sgr~A, but 10$\times$ that of local dense
    clouds.}
  {The ionization rates of interstellar clouds thus seem to be to first order
    determined by the ambient cosmic-ray flux, while propagation effects cause a
    factor of $\sim$3 decrease from diffuse to dense clouds.}
   \keywords{ISM: clouds -- ISM: molecules -- ISM: cosmic rays -- Molecular processes -- Astrochemistry}

   \maketitle
%
%________________________________________________________________

\section{Introduction}
\label{s:intro}

The ionization rate of interstellar clouds is a key parameter for
their dynamics and their chemistry (see
\citealt{caselli-walmsley:ionization} and \citealt{vdtak:london} for
 reviews). In dense star-forming regions, 
the bulk of the ionization is due to cosmic rays, at a rate of
\zcr$\sim$\pow{3}{-17}\,s$^{-1}$ as derived from
\hcop\ and CO emission line observations of the envelopes of young massive
stars up to a few kpc from the Sun \citep{vdtak:zeta}. This result is
in good agreement with measurements of the local flux of low-energy
cosmic rays with the Voyager and Pioneer spacecrafts \citep{webber:voyager}.
However, mm-wave observations of \dcop\ and \hcop\
toward the nearby dark cloud LDN~1544 indicate a decrease of
\zcr\ by a factor $\sim$10 \citep{caselli:zeta}, while
observations of \hhhp\ absorption toward the $\zeta$~Per diffuse cloud
indicate an increase of similar magnitude \citep{lepetit:zeta-per}.
Even larger enhancements are found for the Galactic Center, where
\citet{oka:sgra} derive \zcr$\sim$10$^{-15}$\,s$^{-1}$ from \hhhp\
observations toward Sgr~A. 
An enhanced ionization rate in the inner 250\,pc of our Galaxy is
expected from its strong X-ray and radio synchrotron emission, but
this value is at the upper limit based on the
observed temperatures of the Sgr~B2 clouds \citep{guesten:sgrb2}.
It is unclear if these variations in \zcr\ are due to propagation effects
(absorption or scattering of cosmic rays) or to variations in the
cosmic-ray flux with location in the Galaxy. 
To resolve this issue,
an estimate of the ionization rate of dense gas in the Galactic Center
is urgently needed.
However, in the case of Sgr~B2, using the \dcop/\hcop\ ratio is
  complicated by the uncertain D abundance, while comparing
  \hcop\ with CO is difficult as the molecules may not trace the
  same gas \citep{jacq:sgrb2}, which calls for other methods.

The high proton affinity of water makes hydronium (\hhhop) a key ion
in the oxygen chemistry of dense clouds.
Its submillimeter rotation-inversion transitions may be used to
measure \zcr\ and trace \hho\ and \oo\ which are unobservable from the
ground.
Pioneering work by \citet{wootten:h3o+} and \citet{phillips:h3o+} has
demonstrated this potential
% but was severely limited by atmospheric transmission at Mauna Kea.
with observations of strong \hhhop\ emission toward Sgr~B2 (OH)\footnote{
This position between the Sgr~B2 (M) and (S) cores was used for
  early line surveys of the Galactic Center (e.g.,
  \citealt{cummins:sgrb2oh}) but does not correspond to any object,
  not even the centroid of the OH masers \citep{gaume:oh}. }.
Unfortunately, the atmospheric transmission from Mauna Kea is rarely
good enough to observe \hhhop, so mapping is not feasible.

The Chajnantor site greatly facilitates
observation of the \hhhop\ $J_K$=$3_2^+$--$2_2^-$ line at
364.7974~GHz, which has an upper level energy of 139~K and an
Einstein~$A-$coefficient of \pow{2.8}{-4}~s$^{-1}$. 
%(The 396~GHz line is not accessible to APEX and belongs to the ortho moiety).
The $J_K$=$1_1^-$--$2_1^+$ line ($\nu$=307.1924~GHz, $E_u$=80~K,
$A$=\pow{3.5}{-4}\,\rs) is a probe of denser gas, because the
%its emitting level is the upper level of an inversion pair. The 
emission competes with fast pure inversion decay, unlike for
the 364~GHz line.
This paper presents new observations of these lines
toward the Sgr~B2 region. 
Combined with observations of the
\hhoe\ $3_{13}$--$2_{02}$ line at 203.4075\,GHz ($E_u$=204~K,
$A$=\pow{4.9}{-6}\,\rs), the data are used to estimate \zcr\ in the Sgr~B2 region.
Due to the high critical densities of the lines, our observations are
not sensitive to the extended low-density envelope seen in
absorption lines of water \citep{comito:hdo} and \hhhop\ \citep{goicoechea:h3o+}.
%__________________________________________________________________

\section{Observations}
\label{s:obs}

\begin{table}[tb]
\caption{Measured line parameters at selected positions with
1$\sigma$ errors in units of the last decimal in brackets. Upper limits are
1$\sigma$ limits on \tmb\ in K on 1.0~\kms\ channels.}
\label{t:line}
\begin{tabular}{lllrrr}
\hline
\hline
Position  & $\alpha$(J2000) & $\delta$(J2000) & $\int$\tmb dV & \vlsr & \dv \\
          & hh mm ss & $^0$ $'$ $''$ & K\,\kms         & \kms  & \kms \\
\hline
\multicolumn{6}{c}{\hhhop\ 364 GHz:} \\
Sgr\,B2\,(M)  & 17:47:20.2 & $-$28:23:05 & 42.0(13) & +62.0(6) & 17.3(16) \\  %tmb = 2.1
Sgr\,B2\,(OH) & 17:47:20.8 & $-$28:23:32 &  8.7( 4) & +62.7(3) & 15.3( 8) \\  %tmb = 0.60
%(N)          & $<$0.4   & ...      & ...      \\ just mention confusion limited
GCS 3-2   & 17:46:14.9 & $-$28:49:43 & $<$0.09 & ...      & ...      \\
GC IRS3   & 17:45:39.6 & $-$29:00:24 & $<$0.10 & ...      & ...      \\
\hline
\multicolumn{6}{c}{\hhhop\ 307 GHz:} \\
Sgr\,B2\,(M)      & & & 27.1(52) & +64.6(14) & 16.6(21) \\  %tmb = 1.67
Sgr\,B2\,(OH)     & & &  7.7( 7) & +64.9(13) & 24.0(22) \\  %tmb = 0.33
%(N)
\hline
\multicolumn{6}{c}{\hhoe\ 203 GHz:} \\
Sgr\,B2\,(M)      & & & 20.5(3) & +63.3(1) & 13.2(1) \\  %tmb = 1.47
Sgr\,B2\,(OH)     & & & $<$0.03   & ... & ... \\ 
%SgrB2(N)$^e$
\hline
% \multicolumn{4}{l}{$^a$: $\alpha$=17:47:20.2, $\delta$=$-$28:23:05 (J2000)} \\
% \multicolumn{4}{l}{$^b$: $\alpha$=17:47:20.8, $\delta$=$-$28:23:32 (J2000)} \\
% \multicolumn{4}{l}{$^c$: $\alpha$=17:46:14.9, $\delta$=$-$28:49:43 (J2000)} \\
% \multicolumn{4}{l}{$^d$: $\alpha$=17:45:39.6, $\delta$=$-$29:00:24 (J2000)}
% %\multicolumn{4}{l}{$^d$: $\alpha$=17:47:20.4, $\delta$=$-$28:22:21 (J2000)}
\end{tabular}
\end{table}

In July 2005, we observed the 364~GHz line of \hhhop\ towards selected
sources in the Sgr~B2 and Sgr~A clouds using the APEX telescope
(G\"usten et al., this volume)\footnote{This paper is based on data
  acquired with the Atacama Pathfinder EXperiment, which is a
  collaboration between the Max-Planck-Institut f\"ur Radioastronomie,
  the European Southern Observatory, and the Onsala Space
  Observatory.}. The front end was the facility APEX-2a receiver built
at Onsala (Risacher et al., this volume); the back end was an
FFT spectrometer built at the MPIfR providing 8192 channels over a
bandwidth of 1.0~GHz (Klein et al., this volume). A map of
Sgr~B2 at 10$''$ spacing (Fig.~\ref{f:map}) was made by integrating
1~minute per point, which for system temperatures of 340 -- 360~K gave
a noise level of 0.2~K per 0.8~\kms\ channel. Deeper integrations
(2--5~min) were made at the (N), (M) and (OH) positions.
The line was also observed toward two positions in the Sgr~A cloud,
CGS 3-2 and GC IRS3,
where strong \hhhp\ mid-infrared absorption suggests a high ionization
rate \citep{oka:sgra}.  The non-detections at these positions probably
reflect a lack of dense gas along these lines of sight.
Follow-up observations of the 307~GHz line of \hhhop\ were performed
in November 2005. Since this line lies very close to a CH$_3$OH line,
several local oscillator settings were used to verify the sideband
origin of the detected features.
In this case, the system temperature of 300~K resulted in an
rms noise level of 90~mK on 0.24~\kms\ channels after 10.4 minutes of
on-source integration. 

All data were obtained using position switching with a throw of 3$'$ in Az.
Test spectra taken at the Sgr~B2 (OH) offset position indicate that the
emission is $<$10\% of the strength at the target position.
The beam size of APEX is $\sim$18$''$ at 364~GHz and $\sim$21$''$ at 307~GHz.
Data at both frequencies were calibrated onto \tmb\ scale using
forward and main beam efficiencies of 97\% and 74\%. 
From a comparison of calibrations using hot and cold loads and using a
water vapour radiometer, we estimate the calibration uncertainty to be
$\approx$10\%. 
For reasonably strong lines (\tmb$\gtsim$10$\times$rms), as is the case
here, spectral rms has a negligible contribution to the intensity
uncertainty. 

Observations of the \hhoe\ 203~GHz line were performed with the IRAM 30-m
telescope\footnote{IRAM is an international institute for
  millimeter-wave astronomy, co-funded by the Centre National de la
  Recherche Scientifique (France), the Max-Planck-Gesellschaft
  (Germany) and the Instituto Geografico Nacional (Spain)}
in February 2006. The facility receiver A230 was used as
front end and the VESPA autocorrelator as backend. At this wavelength,
the telescope has a beam size of 12$''$ and a main beam efficiency of
57\%. System temperatures were 500--600~K and integration times
10--30~min, resulting in rms noise levels of $\sim$0.1~K on 0.20~\kms\
channels. The line was detected at the Sgr~B2 (M) but not at the (OH)
position. Calibration was verified to 10\% on Sgr~B2 (N) which was
observed before by \citet{gensheimer:h2o}.

\section{Results}
\label{s:spat}

\begin{figure}[tb]
\centering
\includegraphics[height=9cm,angle=0]{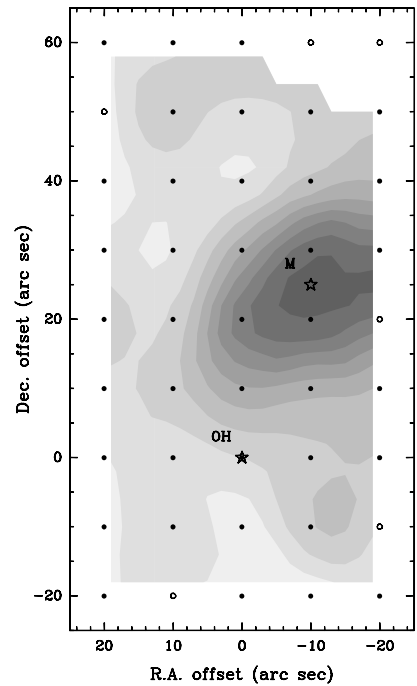}  
\caption{Distribution of velocity-integrated \hhhop\ 364~GHz emission toward
  Sgr~B2, observed with the APEX telescope. Filled dots indicate detections; open dots
  denote upper limits. The greyscale runs from 1.9 to 35.5 in steps of 3.7 \kkms.}
\label{f:map}
\end{figure}

Emission in the \hhhop\ 364~GHz line is detected over an
(80$\times$40)$''$ region in Sgr~B2 (Fig.~\ref{f:map}), corresponding
to (3.3$\times$1.6)~pc at a distance of 8.5~kpc.
The overall distribution of \hhhop\ 364~GHz in the Sgr~B2 region
follows that of many other molecular lines when mapped at comparable
resolution (e.g., \citealt{martin-p:sgrb2}; \citealt{vicente:sgrb2}).
The distribution of \hhhop\ also follows that of the \smm\ dust
continuum emission \citep{lis:sgrb2}, a common tracer of the \hh\ 
distribution.
In contrast, the \hho\ 183.31~GHz emission is strongly peaked at the
cores, dropping to $\ltsim$10\% of its peak value at the (OH) position
\citep{cernicharo:h2o}. This behaviour does not necessarily
reflect the \hho\ column density distribution because of the maser
nature of the 183~GHz line.

Table~\ref{t:line} lists the results of fitting Gaussian profiles to
the spectra after subtracting a polynomial baseline.
The 307~GHz line strength toward OH may be overestimated due to
blending: fixing \dv\ to 15--16\,\kms\ reduces the line flux by
25--30\%.  The strength of this line agrees with the data of
\citet{phillips:h3o+}, while the 364\,GHz line is a factor of 2
weaker. The steep emission gradient (Fig.~\ref{f:map}) makes the
measurement sensitive to pointing offsets. In the APEX data, \hhhop\
and other lines peak at the M core, validating the pointing to
$<$5$''$.

The 307 and 364 GHz spectra show many lines in addition to \hhhop.
While the Sgr~B2 (OH) spectra show $<$10 mostly weak
(\tmb$\ltsim$0.5~K) lines over the 1~GHz bandwidth, the Sgr~B2 (M)
spectrum shows $>$20 features with \tmb$\gtsim$1~K.
The spectra taken in the vicinity of Sgr~B2 (N) are so confused that the \hhhop\ line parameters
cannot be reliably extracted.  Identification and interpretation of
these other lines will be done elsewhere in the framework of a
spectral line survey program (Belloche et al., in prep.).

\begin{figure}[tb]
\includegraphics[width=4.5cm,angle=-90]{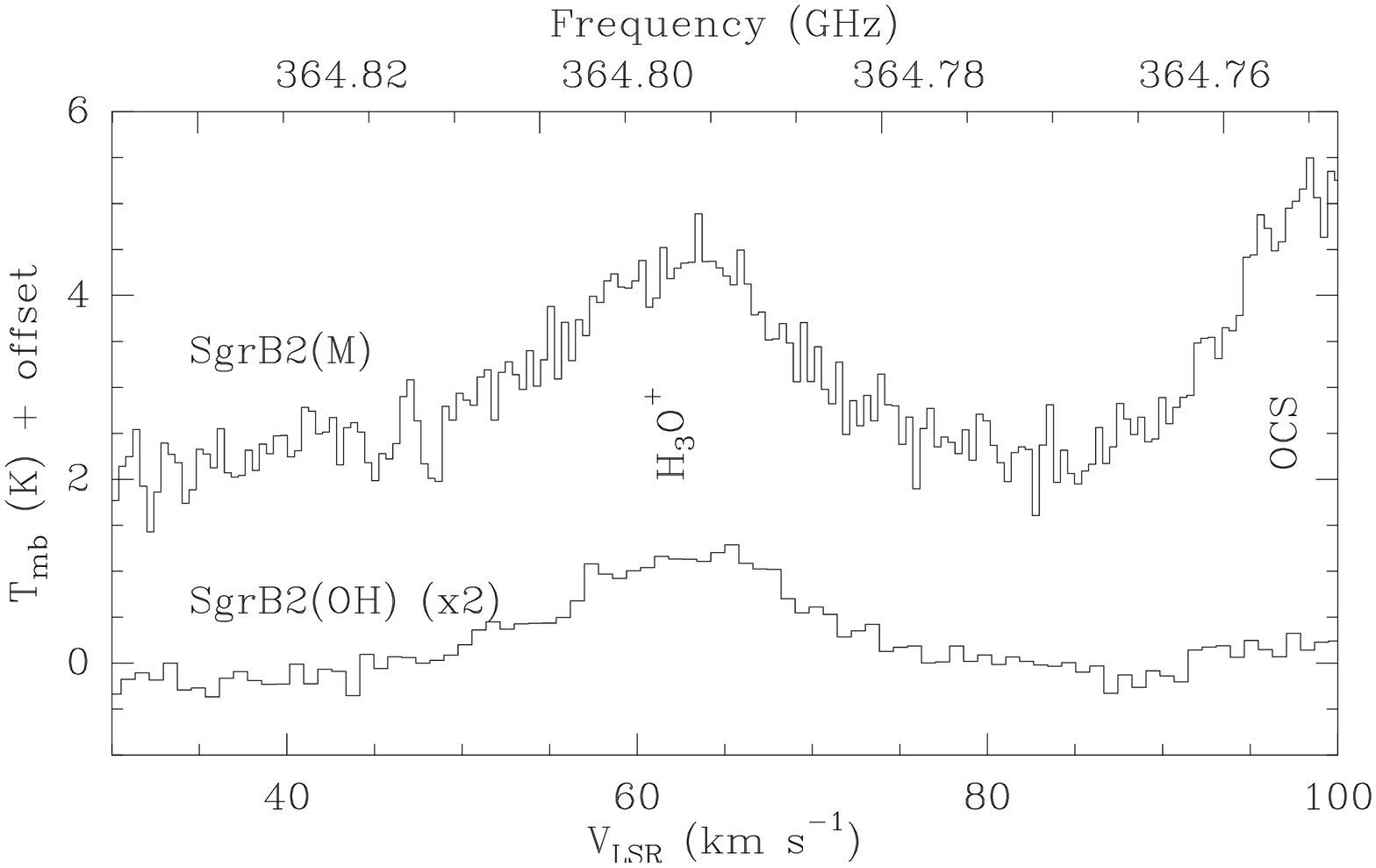}  
\bigskip
\includegraphics[width=4.5cm,angle=-90]{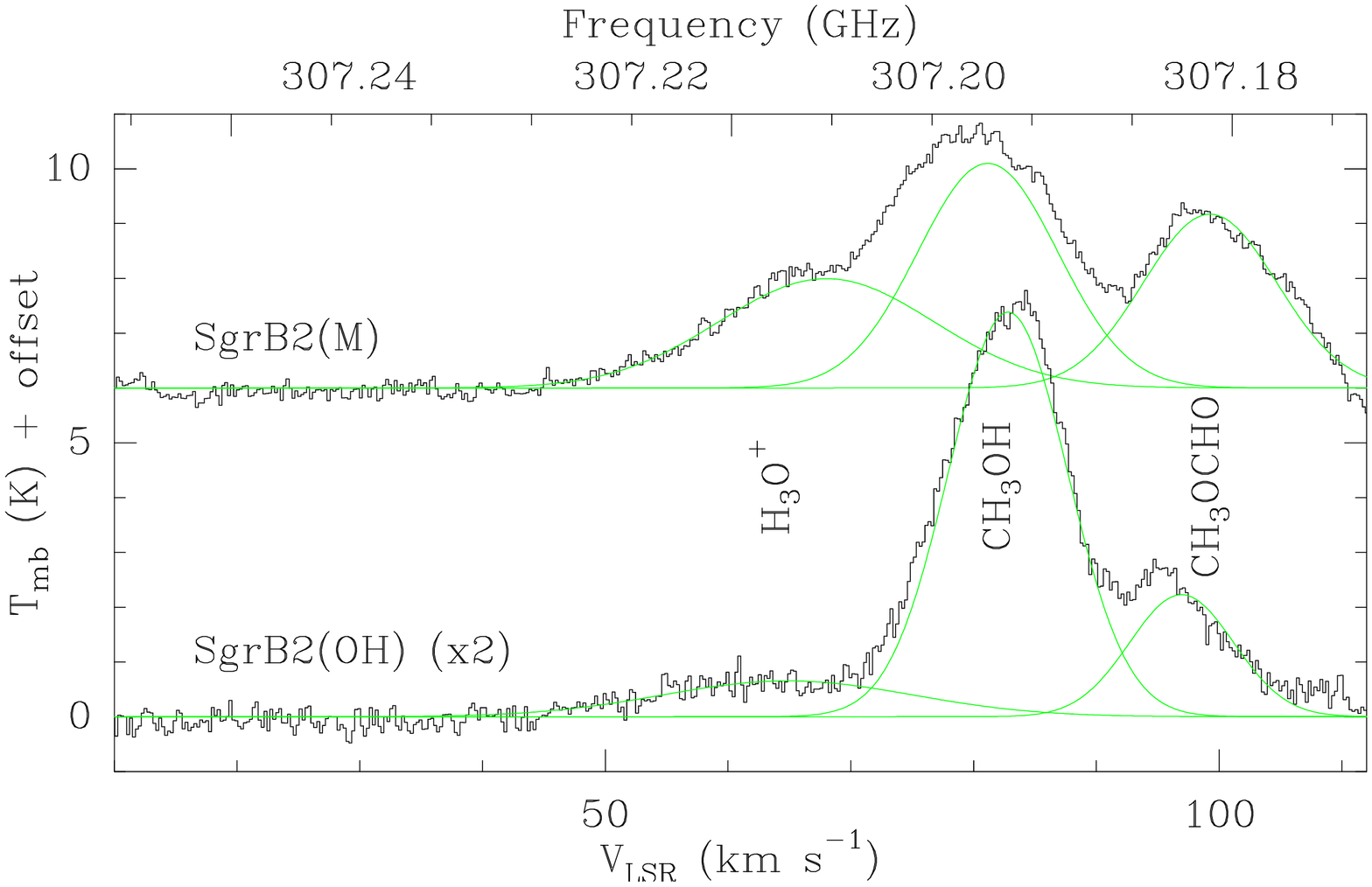}
\vspace{-0.5cm}  
\caption{Spectra of the \hhhop\ 364~GHz line (top) and 307~GHz line
  (bottom) toward the Sgr~B2 (M) and Sgr~B2 (OH) positions.} 
\label{f:spec}
\end{figure}

\section{Column densities of \hhhop\ and \hho}
\label{s:cold}

\begin{table}[tb]
\caption{Column densities of \hho, \hhhop, and \hh\ in the Sgr~B2 region.}
\label{t:chem}
\begin{tabular}{lccc}
\hline
\hline
Component     & $N$(\hho) & $N$(\hhhop)   & $N$(\hh)$^a$   \\
       & 10$^{18}$\,\scm  & 10$^{15}$\,\scm & 10$^{24}$\,\scm \\
\hline
Core (M)      & 35        & 13.6  & 4.4 \\
Envelope (OH) & 0.15$^b$  &  3.7  & 1.1 \\ 
\hline
Beam size ($''$)& 12 & 18 & 20 \\
\hline
\multicolumn{4}{l}{$^a$: From CSO 350\,\mic\ data assuming $T_d$=40\,K} \\
\multicolumn{4}{l}{and $\kappa_\nu$=0.07\,cm$^2$g$^{-1}$ (Lis, priv.\ comm.)} \\
\multicolumn{4}{l}{$^b$: From \citet{cernicharo:h2o} assuming \hho/\hh=10$^{-7}$} \\
\end{tabular}
\end{table}

Figure~\ref{f:spec} shows spectra of the two \hhhop\ lines at the
positions where 307~GHz observations have been made.
At both positions, the intensities of the two \hhhop\ lines are comparable. 
Since the lines are optically thin for any chemically reasonable
\hhhop\ abundance, the line ratio of $\sim$unity indicates a high
excitation temperature, \txc$\gtsim$50~K. 
Statistical equilibrium calculations using molecular data from
\citet{schoeier:moldata}\footnote{\tt
  http://www.strw.leidenuniv.nl/$\sim$moldata/radex.php} 
indicate that at the conditions in the M core ($T$=200\,K,
$n$(\hh)=10$^7$\,\ccm), collisions can sustain \txc$\approx$100\,K,
consistent with the observed \hhhop\ line ratio.
However, the density (10$^6$\,\ccm) at the OH position is too low to
thermalize the excitation at the kinetic temperature of 60\,K. The
observed \hhhop\ line ratio therefore indicates that the \hhhop\ 
excitation is driven by far-infrared pumping by the strong dust
continuum radiation in the Sgr~B2 region, as found before by
\citet{phillips:h3o+}. Models with $T_d$=40--80\,K reproduce the
observed line ratio.
Far-infrared pumping is also expected to play a major role toward
  (M), due to its high infrared luminosity, and is therefore included
  in the model.

For these excitation conditions, the observed line strengths imply
$N$(p-\hhhop)=\pow{1.83($\pm$0.27)}{15}~\scm\ for Sgr~B2 (OH) and
\pow{6.81($\pm$1.01)}{15}~\scm\ for Sgr~B2 (M).
The total \hhhop\ column densities are twice these values, as the
ortho/para (\opr) ratio tends to unity at high temperatures ($\gtsim$100~K).
These estimates have uncertainties of a factor of $\sim$2 due to the
uncertain excitation and the difference in estimated values from
the two lines.
Since the lines are optically thin and the source sizes cannot be
constrained, the derived column densities are beam averages.
For the Sgr~A sources, assuming \dv=10\,\kms, \txc=100\,K and \opr=1,
the observed limit on the line flux of $\approx$1\,\kkms\ corresponds
to $N$(\hhhop) $<$\pow{6}{13}\,\scm.

Dividing the above column density estimates by $N$(\hh) values based
on \smm\ dust continuum mapping (Table~\ref{t:chem}) provides
estimates of the column-averaged abundance in the two sources:
$N$(\hhhop)/$N$(\hh) = \pow{3.1}{-9} in Sgr~B2 (M) and \pow{3.4}{-9}
in Sgr~B2 (OH).
These estimates of the \hhhop\ column density and abundance are
higher than previous values by
\citet{phillips:h3o+} which were derived from the \hhhop\
  396\,GHz line assuming a density of 10$^5$\,\ccm\ for the Sgr~B2
  envelope, a value which likely is an order of magnitude too low
  \citep{goicoechea:iso-lws}.

To estimate the \hho\ column density from the IRAM observations, we
adopt the same excitation conditions as for \hhhop. To convert
p-\hhoe\ into \hho, we adopt \opr=3 as applicable for warm gas and 
$^{16}$O/$^{18}$O=250 as applicable for the Galactic Center
\citep{wilson:abundances}.
For the Sgr~B2 (M) position, we estimate $N$(\hho)=\pow{3.5}{19}\,\scm,
while the upper limit at the (OH) position implies
$N$(\hho)$<$\pow{1.7}{18}\,\scm\ (1$\sigma$). Within the uncertainty of
a factor of two, our estimate of $N$(\hho) at the M position is
consistent with radiative transfer modelling of the \hhoe\ 203\,GHz
line towards the N core \citep{comito:hdo}.
The \hho\ column densities imply abundances of \pow{1.2}{-5} and
$<$\pow{1.7}{-6} at the (M) and (OH) positions, respectively, consistent
with the estimates of $\sim$10$^{-5}$ and $\sim$10$^{-7}$ from
observations of the \hho\ 183~GHz line \citep{cernicharo:h2o}.

\section{Chemistry of \hho\ and \hhhop}
\label{s:disc}

Table~\ref{t:chem} summarizes measurements of the \hho\ and \hhhop\
column densities toward the Sgr~B2 cores (represented by the M
position) and the surrounding envelope (represented by the OH
position).
The \hhhop/\hho\ ratio is seen to be $\sim$1/50 in the envelope,
or about 20 times the value in local clouds
\citep{phillips:h3o+}, but drops by $\sim$100 toward the cores.
This drop is more than expected from enhanced recombination at higher
densities, and may be due to ongoing injection of \hho\ into the gas
phase by evaporation of icy grain mantles, which also causes the rich
\smm\ spectra of the (M) and (N) cores.

To estimate the cosmic-ray ionization rate \zcr\ in the Sgr~B2 region from
our observations, we have used the chemical model of \citet{vdtak:zeta}.
Given estimates of temperature, density, \zcr, and the abundances of
CO, O, \nn\ and \hho, this model calculates the steady-state
abundances of \hcop, \nnhp\ and \hhhop, as well as the total electron
fraction.
The model neglects metals such as Mg, Fe and S as electron sources and
PAHs as electron sinks; including these species may change the
calculated electron fraction by 2--3 depending on their abundance.
The reaction rates in the model have been updated to follow the UMIST99 values
\citep{leteuff:umist99}, except for the dissociative recombination
reactions of \hhhp, \hcop\ and \nnhp\ where the values of \citet{mccall:zeta_per},
\citet{boisanger:ions} and \citet{geppert:n2h+} are used.
For the cores, we adopt $T$ = 200~K and $n$ = 10$^7$ \ccm, while for the 
envelope, we use $T$ = 60~K and $n$ = 10$^6$ \ccm\ (see \citealt{goicoechea:iso-lws}).
The CO abundance of \pow{5}{-5} is from the SEST line survey \citep{nummelin:sgrb2},
the O abundance of \pow{1.5}{-4} is based on the ISO-LWS line survey \citep{goicoechea:iso-lws},
and we assume equal CO and \nn\ abundances.

The model reproduces the observed \hhhop/\hho\ ratio in the Sgr~B2 envelope
for \zcr$\sim$\pow{4}{-16}\,\rs. The uncertainty is a factor of 2 from
the observed \hhhop/\hho\ ratio, and another factor of 2 through
the assumed CO, \nn\ and O abundances. 
%
% The model also reproduces the observed \hcop\
% abundance of \pow{2}{-10} \citep{nummelin:sgrb2}. 
% %
% The abundance of
% \nnhp\ is predicted to be similar to that of \hcop, but is not
% constrained by observations. 
%
For Sgr~B2 (M), the model indicates a $\sim$10$\times$ lower \zcr,
which we consider a lower limit because the chemistry may not be in
steady state. 
The time scale to equilibrate the \hho/\hhhop\ ratio is $t_{\rm ion} \sim
30(n_e)^{-0.5}$~yr \citep{vdtak:h2o} which for the model electron
density of $n_e$=0.01~\ccm\ equals 300~yr.

The derived value of \zcr\ for Sgr~B2 (OH) is 10$\times$ higher than the
value for dense gas in the Solar neighbourhood (\S~1), but 3$\times$
lower than indicated by \hhhp\ observations of the Sgr~A region
\citep{oka:sgra}.
%
% The fact that the Sgr~B2 clouds are the most massive molecular clouds
% in the Galaxy suggests that the ionization rate of dense clouds is
% mostly determined by their column density, whereas local variations in
% the cosmic-ray flux play a minor role.
% %
% The penetration depth of cosmic rays is $\approx$100~g\,\scm\
% \citep{umebayashi:cosmic-rays}, which corresponds to
% $N$(\hh)$\sim$10$^{25}$\,\scm. The Sgr~B2 clouds have somewhat smaller
% column densities, so that absorption of cosmic rays probably plays a
% minor role. The apparent scaling of the ionization rate with density
% may instead be due to cosmic-ray scattering off plasma waves. This
% process is more effective in denser clouds where magnetic fields are
% stronger \citep{padoan:cosmic-rays}, which would explain the observed
% trend. 
We conclude that the ionization rates of dense molecular clouds are
mainly determined by their location in the Galaxy through variations in the
cosmic-ray flux by a factor of $\sim$10. As a second order effect,
\zcr\ is 3$\times$ lower in dense molecular clouds than in diffuse clouds, presumably due
to cosmic-ray scattering \citep{padoan:cosmic-rays}.

In the future, large-scale mapping of the Sagittarius molecular clouds
in the \hhhop\ 364\,GHz and \hhoe\ 203\,GHz lines, combined with
constraints on the temperature and density structure of the clouds,
will provide more detailed understanding of the nature of the
interaction between cosmic rays and molecular gas. The IRAM and APEX
telescopes are well suited to carry out such observations, especially
if equipped with heterodyne array receivers.
Observations of \hho\ and \hhhop\ ground-state lines with the
\textit{Heterodyne Instrument for the Far-Infrared} (HIFI) on the
\textit{Herschel Space Observatory} will allow us to extend these
studies to regions of lower column density.

\begin{acknowledgements}
  The authors thank the APEX staff for making the \hhhop\ observations
  possible, Jos\'e Cernicharo and Juan Pardo for communicating their
  183\,GHz results in advance of publication, Clemens Thum, Nuria
  Marcelino and St\'ephane Leon for arranging the \hhoe\ observations
  on short notice, and Darek Lis and Malcolm Walmsley for useful
    comments on the manuscript. 
\end{acknowledgements}

\bibliographystyle{aa}
\bibliography{zeta}

\end{document}